# Electronic, optical and thermoelectric properties of sodium pnictogen chalcogenides: A first principles study

I. S. Khare[a,b], N. J. Szymanski[c], D. Gall[d], R. E. Irving[b],*

[a]Ottawa Hills High School, Ottawa Hills, OH 43606, USA

[b]Department of Physics and Astronomy, University of Toledo, Toledo, OH 43606, USA

[c]Department of Materials Science and Engineering, University of California, Berkeley, CA 94720, USA

[d]Department of Materials Science and Engineering, Rensselaer Polytechnic Institute, Troy, NY 12180, USA

*Corresponding Author: richard.irving@utoledo.edu

## Abstract

Ternary chalcogenides have been of recent investigation for applications in photovoltaic and thermoelectric devices. We study the structural, electronic, optical, and thermoelectric properties of nine ternary chalcogenides, $NaAX_2$, where A represents pnictogens (As, Sb, and Te) and X represents chalcogens (S, Se, and Te). Calculations based on density functional theory yield the following results: (i) phonon dispersion curves predict three of the compounds, $NaAsS_2$, $NaSbS_2$, and $NaSbSe_2$, to be dynamically stable in the monoclinic, *C2/c*, structure, (ii) the layered atomistic configuration causes the corresponding electronic and optical properties to display a high degree of anisotropy, (iii) A-X electronic bonding features vary significantly with structural distortions arising from atomic size mismatch, therefore directly influencing stability, (iv) strong absorption is observed in the stable compounds, with coefficients ranging from $10^4$ to $10^5$ cm-1 in the visible-UV range, and (v) remarkably high Seebeck coefficients exceeding 500 $\mu$V/K at carrier concentrations commonly achieved in such materials. From these results, we conclude that $NaAsS_2$, $NaSbS_2$, and $NaSbSe_2$ are suitable candidates for both photovoltaic, particularly in tandem solar cells, and thermoelectric applications. Experimental synthesis and verification are suggested.

## 1. Introduction

There exists a rising need for renewable energy and greater energy efficiencies due in part to the limited supply of non-renewable energy sources such as fossil fuels [1]. New technological advances are required to achieve the necessary growth, particularly in the areas of photovoltaics (PV) and thermoelectrics. Beyond silicon, material classes such as cadmium telluride, gallium arsenide, copper indium diselenide, perovskites, nitrides, oxides, and organic materials have been commonly studied for solar energy conversion [2-9]. Previously studied material classes for thermoelectric (TE) applications include binary chalcogenides such as $Bi_2Te_3$, lead telluride, inorganic clathrates including both $A_xB_yC_{46-y}$ (type I) and $A_xB_yC_{136-y}$ (type II), and oxides [10-14].

More recently, ternary chalcogenides have been of interest for both photovoltaic and thermoelectric applications [15-20]. Such chalcogenides, including $NaSbS_2$ [21], $KSbS_2$ [22], $RbSbS_2$ [23], $KSbSe_2$ [23], $NaAsSe_2$ [24], $NaSbSe_2$ [24], $NaSbTe_2$ [24], and $NaBiTe_2$ [24] have been experimentally synthesized in various structures, and were found to display interesting properties [24-26]. $NaSbS_2$, which is stable in the *C2/c* space group, exhibits a reasonably large optical absorption coefficient, $\alpha$, ranging from $10_4 - 10_5$ cm$_{-1}$ in the visible-UV range [25]. The first $NaSbS_2$ semiconductor-sensitized solar cells were demonstrated to have an efficiency of $\eta = 3.18\%$ under 0.1 suns which is respectable for an emerging PV material [27]. Furthermore, when alloyed with another ternary chalcogenide in the *C2/c* structure, $NaSbSe_2$, the material was predicted to possess an optimal band gap of 1.4 eV [18]. Earlier works have also suggested these two compounds to be appealing for thermoelectric applications [17, 20]. $NaSbS_2$ has been highlighted specifically for its unique electronic band structure, showing regions of the Brillouin zone with both high and low dispersion of strong character, leading to multi-valley features along the conduction band minimum which are thought to contribute to optimal thermoelectric behavior. These properties, along with substantial acoustic-optical phonon mixing which beneficially influences thermal conductivity, suggest $NaSbS_2$ to be a promising candidate for application in thermoelectric devices [17].

Given the advantageous properties of the known ternary chalcogenides adopting the *C2/c* structure, it would be beneficial to expand the number of such compounds available for PV and TE applications. Therefore, we have studied the structural, electronic, thermoelectric, and optical properties of nine sodium ternary chalcogenides in the monoclinic *C2/c* structure. The compounds are of the form $NaAX_2$, where A represents group 15 pnictogens (As, Sb, and Te), and X represents group 16 chalcogenides (S, Se, and Te). We have implemented first principles methods to study the stability of these materials, as well as their electronic, thermoelectric, and optical properties. By doing so, we have identified several chalcogenides which are promising for use in solar cell and thermoelectric applications.

## 2. Computational Methods

All density functional theory (DFT) calculations have been performed using the Vienna *Ab initio* Simulation Package (VASP) [28-31]. We have employed the Perdew–Burke–Ernzerhof (PBE) exchange correlation functional [32] based on the generalized gradient approximation (GGA) within the framework of the projector augmented wave method (PAW) [33, 34]. A cutoff energy of 600 eV was used for the plane-wave basis set. In addition to outer-core 3s electrons, semi-core 2p electrons were treated as valence states in Na. In bismuth, outer-shell 6s and 6p electrons, as well as semi-core 5d electrons were considered as valence states. Only outer-shell electrons were considered as valence states in As, Sb, S, Se, and Te. All k-point meshes consisted of 4000 k-points per reciprocal atom (KPPRA) [35-40]. A convergence criterion of $10_{-8}$ eV per atom with a Gaussian smearing width of 0.2 eV was chosen for electronic minimizations.

The structure, belonging to the *C2/c* space group, was obtained from experimental observations reported by the literature and used as an initial guess for the NaSbS$_2$ structure [21]. It was then geometrically optimized, allowing the unit cell shape, volume, and ionic positions to be relaxed until the forces acting on each atomic coordinate were less than or equal to 1 meV/Å [41-45]. The remaining seven compounds in this study (NaAsS$_2$, NaAsSe$_2$, NaAsTe$_2$, NaSbTe$_2$, NaBiS$_2$, NaBiSe$_2$, and NaBiTe$_2$) have not been experimentally synthesized in the structure of interest, and therefore no structural data could be directly utilized. Instead, we constructed a unit cell with dimensions and ionic positions equal to that of the known chalcogenide and used this as a starting point for the relaxations [46]. The primitive vectors of the unit cell are given by the following equations: $a_1 = \frac{1}{2}(a\hat{x} - b\hat{y})$, $a_2 = \frac{1}{2}(a\hat{x} + b\hat{y})$, and $a_3 = c\,(\cos\beta\,\hat{x} + \sin\beta\,\hat{z})$ [47, 48]. The general unit cell for this monoclinic structure is displayed in Fig. 1. Optimization of each structure was performed until convergence was obtained, in the same manner as previously described.

To investigate the dynamical stability of these chalcogenide compounds, we have computed phonon dispersion curves using the harmonic approximation [49, 50], which works well under the assumption that anharmonic effects are insignificant. By employing zero-temperature DFT computations, we calculated the Hessian matrix of each compound for $2 \times 2 \times 2$ supercells using density functional perturbation theory (DFPT) [28-31]. Utilizing these results along with the PHONOPY code, [51] harmonic phonon frequencies were calculated along high-symmetry paths in the Brillouin zone.

To improve upon the well-known shortcoming of standard DFT with respect to accuracy of band gap calculations owing to correlation and self-interaction errors, the Heyd-Scuseria-Ernzerhof hybrid functional (HSE06) [52] was employed to study all electronic and optical properties. Hybrid functionals incorporate a portion of exact exchange from Hartree–Fock theory, mixed with exchange from standard DFT methods. For the HSE06 functional, 25% of the Hartree–Fock exact exchange is used, while the remaining 75% is from GGA. In addition to the electronic density of states, Crystal Orbital Hamilton Populations (COHP) were calculated using the LOBSTER software package [53-57].

To study the optical performance of these materials for PV, the complex dielectric function ($\varepsilon_1$ + $i\varepsilon_2$) [58-60] was tabulated as a function of photon energy, E = $\hbar\omega$, for each material. Extinction coefficients $\kappa(\omega)$ and absorption coefficients $\alpha(\omega)$ were calculated using the following equations [61, 62]

$$\alpha(\omega) = 2\omega\kappa(\omega)/c \text{ where } \kappa(\omega) = [((\varepsilon_1{}^2 + \varepsilon_2{}^2)^{1/2} - \varepsilon_1)/2]^{1/2}, \qquad (1)$$

and $c$ is the speed of light.

Thermoelectric properties, namely, the Seebeck coefficient ($S_C$) and reduced power factor (RPF), were found using the Boltzmann Transport Properties (BoltzTraP) program [63], in which a smoothed Fourier interpolation of the bands is used to calculate the derivatives necessary to obtain transport distributions. To calculate the RPF, which is more directly reflective of thermoelectric performance, we used the equation

$$RPF = S_C{}^2\sigma/\tau, \qquad (2)$$

where $\sigma$ is the electrical conductivity, and $\tau$ is the relaxation time [20, 64]. These properties were computed using carrier concentrations from $10_{16}$ to $10_{22}$ cm-3, which have been shown to be achievable by doping [65-67].

## 3. Results

### 3.1 Structure and stability

We find that all nine compounds retain the _C2/c_ structure after full geometric optimization. Optimized lattice parameters for these compounds are displayed in Table 1. The _C2/c_ structure, being monoclinic, has three independent lattice constants _a, b, c_, and one angle $\beta$. For these nine chalcogenides, _a_ ranges from 7.78 to 8.80 Å, _b_ ranges from 7.95 to 9.09 Å, and _c_ ranges from 6.47 to 7.79 Å. Our calculated values match well with both experimental and theoretical data where it is available from previous works [20, 21], with deviations on the order of 1-2% being typical for such comparisons [68]. All lattice constants, and therefore the volume of the unit cell, are found to most strongly correlate with the ionic radius of X owing to its large size. In contrast, the ionic radius of A strongly influences the _c-_ axis length as caused by contraction or expansion of the strong A-X bonds occurring along the direction of *c* [69].

Calculated phonon dispersion curves are shown in Fig. 2. We predict three compounds, $NaAsS_2$, $NaSbS_2$, and $NaSbSe_2$, to be dynamically stable as their phonon dispersion curves are completely real-valued (i.e., no unstable imaginary phonon frequencies are observed). This is consistent with the successful experimental synthesis of both $NaSbS_2$ and $NaSbSe_2$ as reported in the *C2/c* structure [21, 27, 70]. For the yet to be synthesized $NaAsS_2$, our phonon calculations represent a novel prediction of stability. Each of these three compounds are found to exhibit relatively flat acoustic phonon modes in the dispersion curves, as well as mixing of these modes with low-frequency optical branches. Such mixing has been linked to being beneficial for low lattice thermal conductivities [71, 72].

We observe that the ratio of the van der Waals radius of A to the van der Waals radius of X provides an indicator of stability throughout the majority of compounds [73, 74]. Specifically, if the ratio of the van der Waals radii A/X is greater than 1, the compound is stable. The only exceptions to this observation are these containing Bi, which can likely be attributed to the fact that Bi has a large radius which may interact unfavorably with Na. Moreover, the low electronegativity of bismuth leads to more highly ionic bonds, whereas we speculate that this structure prefers more highly covalent A-X bonds. Of the other unstable compounds, $NaAsSe_2$ and $NaSbTe_2$ have been synthesized in the orthorhombic and NaCl-type structures respectively [24]. Thus, we reason that the larger mismatch between A and X leads to stability in the *C2/c* structure.

To further illustrate the deviations from ideal octahedral geometry occurring in the structure of these compounds, bond length and angle distributions are plotted in Fig. 3. The bond lengths of each compound are shown to be separated into groups distinguished by the varying cation-anion pairs. The median bond length is dominated by the long A-X bonds as mentioned earlier. With respect to the bond angles, deviations from ideality are found to most strongly correlate with the size of X, which further reflects the importance of the size mismatch effect influencing this structure. As for compounds containing Bi, we find much less deviation in the bond angles, suggesting these compounds to prefer a simple rocksalt-type structure, whereas in the remaining compounds containing As and Sb, the local environment of the A atom is characterized by elongation of two bonds relative to the remaining four, therefore distinguishing it from the usual octahedral geometry. Such features are the main distinguishing characteristic of the stable *C2/c* structures in contrast to the rocksalt structure.

## 3.2 Electronic properties

Projected electronic densities of states (PDOS) for each compound are displayed in Fig. S1 of the supplementary material and specifically for the dynamically stable ones in Fig. 4. We find that for NaAX$_2$, the clear majority of X states (S, Se, Te) are occupied, i.e., lie below the Fermi energy. There are about an equal proportion of A (As, Sb, Bi) states that are occupied as unoccupied. Lastly, the Na states are found both above and below the Fermi energy, with the majority of states unoccupied. These features agree with results from previous computational works for NaSbS$_2$ and NaSbSe$_2$ [17, 18]. The electronic band gaps for all nine compounds, obtained from HSE DOS, can be found in Table 2. Our values match well with both experimental data and previously calculated values where available [18-20, 27]. We observe that the band gaps for all compounds range from 0.5 to 2.0 eV, and from 1.0 to 2.0 eV for the stable compounds, with the magnitude showing clear correlation with the degree of ionicity between A and X. The Shockley and Queisser limit implies that solar energy conversion with greater than 30% efficiency is possible for semiconductors with band gaps of 0.8–1.7 eV [75]. Thus, based solely upon their electronic properties, six of the nine compounds are potential candidates for photovoltaic applications.

We also plotted the molecular orbital projected density of states (not shown) for the stable compounds, NaAsS$_2$, NaSbS$_2$, and NaSbSe$_2$, to gain more insight into the electronic properties of the compounds. These three compounds should be regarded as nominally ionic (with respect to Na interaction) as the valence bands are derived from the S $p$ and Se $p$ states [19]. Moreover, the top of the valence band is observed to split off to higher energy as is observed in some S compounds due to the repulsion between a lower-lying metal state and the S $p$ states [76, 77]. We infer that the same repulsion exists in Se compounds because the same valence band split to higher energy is observed in NaSbSe$_2$. Additionally, at the top of the valence bands, including at the valence band maximum (VBM), there is As $s$ and Sb $s$ character. Besides this, there is also a considerable amount of As $4p$ and Sb $5p$ character in the valence bands. These states are often unoccupied in these compounds and form the main conduction bands. Consequently, the As $4p$ and Sb $5p$ contributions in the valence bands derive from cross-gap hybridization between the occupied X $p$ states and the unoccupied A $p$ states. In NaSbSe$_2$, we observe that the bottom portion of the conduction band is mainly composed of the anti-bonding states of the Sb $5p$ and Se $4p$ hybridization. In contrast, the top part of the valence band is mainly composed of Sb $5s$, $5p$, and Se $4p$ states. Thus, the VBM state has antibonding character of the Sb $5s$ and Se $4p$ hybridization. A similar feature can be observed in both NaAsS$_2$ and NaSbS$_2$, where the bottom of the conduction band is mostly anti-bonding states of As $4p$ and S $3p$ hybridization and Sb $5p$ and S $3p$ hybridization respectively. The top part of the valence band is also mainly composed of As $4s$, $4p$, and S $3p$ states for NaAsS$_2$, and of Sb $5s$, $5p$, and S $3p$ states for NaSbS$_2$. As discussed in Ref. [18], this is observed to be distinct from conventional zinc blende semiconductors such as CdTe and CuInSe$_2$ with $s$-$p$ bandgaps with bonding state character for VBM and anti-bonding state character for the conduction band minimum.

To further support these observations, we have computed the Crystal Orbital Hamiltonian Populations (COHP) using the LOBSTER program [54-57]. This method allows us to characterize the bonding and antibonding nature of the individual electronic bonds explicitly. COHP results can be found in Fig. 4 and in Fig. S2 of the supplementary material. For all NaAX$_2$ compounds, we find that the A-X covalent interaction results in strong bonding states at lower energies (-6 to -1 eV) and relatively weaker antibonding states at higher energies (-1 to 0 eV). In contrast, the Na-X interactions consist completely of bonding states below the Fermi energy for all compounds because the bond is almost purely ionic. Filled states near the Fermi energy are dominated by pnictogen-chalcogen bonds of antibonding character. Above the Fermi energy, all A-X states are antibonding, while some Na-X states are mostly antibonding, with some very minute bonding states. In all the As and Sb containing compounds, all other states are negligible; however, in all Bi containing compounds, the Bi-Bi states are moderately significant. These states are bonding at low energies (-6 to -4 eV) and antibonding at higher energies (-4 to 0 eV). Having mostly antibonding states, the NaBiX$_2$ compounds turn out to be unstable as was determined earlier in the discussion on phonons in section 3.1. This feature helps to explain the deviance of bismuth compounds from the observation, as discussed earlier in this section, that a van der Waal radius ratio greater than 1 for A and X is associated with stability.

Tabulated in Table 2 are the negative Integrated Projected Crystal Orbital Hamilton Populations (-IpCOHP), for the cation-anion bonds in each compound. The negative of the energy integration up to the Fermi level yields -IpCOHP, which serves as a bond-weighting indicator [56, 78]. This property can also be potentially useful in explaining the stability of structures. In our compounds, the A-X bonding states vary significantly with structure and therefore influence stability. We observe that the three stable compounds, NaAsS$_2$, NaSbS$_2$, and NaSbSe$_2$, all have the largest -IpCOHP values for the A-X states as observed in Fig. 4 and Table 2. These values are 3.39, 3.47, and 3.39 eV, respectively. The next highest value is 3.23 eV for NaAsSe$_2$ and NaAsTe$_2$. The lowest value of 2.72 eV is for NaBiTe$_2$. We also observe that the stable compounds have the greatest values for the difference between the -IpCOHP of the A-X states and the Na-X states. For each group of compounds, NaAsX$_2$, NaSbX$_2$, and NaBiX$_2$, the -IpCOHP value for the Na-X states is greatest for the Na-Te state. This is also possibly related with the observation, as displayed in Fig. 2, that the NaATe$_2$ compounds have the greatest bond lengths when compared to other compounds of the same group.

Throughout the remainder of this work, we focus on the three compounds which are predicted to be dynamically stable: $NaAsS_2$, $NaSbS_2$, and $NaSbSe_2$. While these compounds may not necessarily be thermodynamic ground states, they are at least metastable, agreeing with data for two known compounds, $NaSbS_2$ [21] and $NaSbSe_2$ [70], which have been synthesized in the *C2/c* structure. Moreover, it has been reported that these compounds are relatively robust with respect to decomposition or polymorphic phase transformation [17, 18, 21, 27, 70].

### 3.3 Thermoelectric properties

The performance of thermoelectric materials is given by the figure of merit (ZT) which is influenced by the electrical conductivity $\sigma$, the Seebeck coefficient $S_C$, temperature, and the thermal conductivity $\kappa$ [79]. However, a high ZT simultaneously requires a high electrical conductivity and a high $S_C$. Compounds of the I-V-VI$_2$ structure are predicted to have very low thermal conductivities as determined by earlier work [80]. Thus, we focus on the electronic aspects of ZT. One of the main challenges in finding high performance TE materials is that as $S_C$ increases, $\sigma$ decreases. Increasing the carrier concentration generally increases $\sigma$, but also reduces $S_C$. The temperature at which the TE device is used also determines its figure of merit. Higher temperatures often result in higher values for both $S_C$ and ZT. Calculated Seebeck coefficients as a function of carrier density for both p-type and n-type doping concentration of all compounds are displayed in Fig. S3 of the supplementary material and in Fig. 5 for $NaAsS_2$, $NaSbS_2$, and $NaSbSe_2$. As in Ref. [20], we calculated these values at three distinct temperatures, 300 K, 450 K, and 650 K, which span the temperatures at which TE devices are commonly used. The higher temperatures are also of particular interest given that a large temperature difference can produce a higher conversion efficiency [81].

We find that at a carrier density of $10_{19}$ cm-3 and a temperature of 300 K, the Seebeck coefficient of $NaSbS_2$ is 200 $\mu$V/K for the p-type and 100 $\mu$V/K for the n-type. At a higher temperature of 600 K, however, the n-type has a larger value of 300 $\mu$V/K, while the value for the p-type is about 250 $\mu$V/K. Between carrier concentrations of $10_{19}$ and $10_{20}$ cm-3 for $NaSbSe_2$ at a temperature of 300 K, the n-type is greatest with $S_C = 125$ $\mu$V/K and the p-type is greatest with $S_C = 150$ $\mu$V/K. At higher temperatures of 450 K and 600 K, the $S_C$ of the p-type is still greater. Between carrier densities of $10_{19}$ to $10_{20}$ cm-3 for $NaAsS_2$, we find the $S_C$ of the n-type to be about equal to the $S_C$ of the p-type. Among the three stable compounds studied, $NaAsS_2$ is observed to have the highest $S_C$ at 600 K. It is about 350 $\mu$V/K for both p-type and n-type. As indicated in Ref. [82], $S_C$ is higher at lower carrier concentrations of $10_{18}$ cm-3. At this concentration, both p-type and n-type $NaAsS_2$ have a very high $S_C$ of over 500 $\mu$V/K at high temperatures. Additionally, n-type $NaSbS_2$ and $NaSbSe_2$ at 600 K have quite remarkable Seebeck coefficients of 500 and 300 $\mu$V/K. These compounds have comparable or even higher $S_C$ values when compared with common thermoelectric materials currently implemented. At a temperature of 300 K and a carrier concentration of $10_{19}$ cm-3, p-type PbTe has a $S_C = 175$ $\mu$V/K [83]. At these same conditions, $NaSbS_2$ and $NaSbSe_2$ have comparable Seebeck coefficients of $S_C \approx 175$ $\mu$V/K while $NaAsS_2$ has $S_C = 250$ $\mu$V/K. When compared with n-type $Bi_2Te_3$, which has a $S_C$ of 200 $\mu$V/K at 300 K, $NaAsS_2$ has a greater $S_C$ of 250 $\mu$V/K [84]. Therefore, we suggest all three stable compounds as potential candidates for TE applications. Of these three, past literature has made similar predictions for $NaSbS_2$ and $NaSbSe_2$ [17-20, 27, 82], but $NaAsS_2$ is a unique prediction from our work.

The reduced power factor (RPF) defined in Section 2 captures the combined effect of both σ and $S_C$. The RPF for $NaAsS_2$, $NaSbS_2$, and $NaSbSe_2$ are plotted in Fig. 5 and in Fig. S4 of the supplementary material for all compounds. Although $NaAsS_2$ has a greater Seebeck coefficient, $NaSbS_2$ and $NaSbSe_2$ have higher reduced power factors as a result of the electrical conductivities. In fact, at a carrier concentration of $10_{20}$ cm-3 and 600 K, $NaAsS_2$ has an RPF of $30 \times 10_{10}$ W/m-1K-2s-1 for the n-type and an RPF of slightly over $10 \times 10_{10}$ W/m-1K-2s-1 for the p-type. At these same conditions, $NaSbS_2$ has RPFs of $5 \times 10_{10}$ and $70 \times 10_{10}$ W/m-1K-2s-1 for the p-type and n-type respectively. The RPF of the n-type for $NaSbS_2$ is over twice that of $NaAsS_2$. Finally, at carrier density $10_{20}$ cm-3 and 600 K, $NaSbSe_2$ has an RPF of just over $5 \times 10_{10}$ W/m-1K-2s-1 for the p-type and $55 \times 10_{10}$ W/m-1K-2s-1 for the n-type. As has been commented on earlier, $NaSbS_2$ and $NaSbSe_2$ may be potentially useful TE materials [17, 19, 20]. Now we note that recently the lattice thermal conductivities of $NaSbS_2$ and $NaSbSe_2$ were theoretically determined to range between 0.753 and 1.173 Wm-1K-1 at 300 K [82]. These values are comparable to those of other thermoelectric materials. For example, a single crystal of PbTe has a thermal conductivity of around 2.2 Wm-1K-1 at 300 K [85], while recently the lattice thermal conductivity of 1% CuI-Pb co-doped $Bi_2Te_3$ was found to be 0.66 Wm-1K-1 at 300 K [86].

## 3.4 Optical properties

Calculated complex dielectric functions are displayed in Fig. S5 of the supplementary material. Due to the monoclinic layered structure of the compounds, the optical properties are highly anisotropic. Accordingly, we have plotted both $\varepsilon_1$ and $\varepsilon_2$ in their directional components. Compounds that contain S, $NaAsS_2$ and $NaSbS_2$, have low values for the static dielectric constants ($\varepsilon_1$ in the range of 5 to 15 at low energies). We also observe that the imaginary portion of the dielectric function, $\varepsilon_2$, remains nearly zero-valued below the band gap for the semiconductors and rises substantially at energies immediately above the gap.

We have also calculated the energy-dependent absorption coefficient, $\alpha(E)$, of each compound to assist in determining which compounds have potential for application in solar cells. These results can be found in Fig. S6 of the supplementary material and in Fig. 6 for the stable compounds NaAsS$_2$, NaSbS$_2$, and NaSbSe$_2$. For all three compounds, the absorption coefficient at higher energy levels (4 to 6 eV) is greatest in the *c* direction, followed by the *a* direction, and the smallest in the *b* direction. As discussed by Ref. [19], this absorption spectrum is similar for both in-plane polarizations in the visible range but differs strongly for the *c* component, which has electric field polarization perpendicular to the NaAX$_2$ sheets. In the visible range, all three compounds, NaAsS$_2$, NaSbS$_2$, and NaSbSe$_2$, have large peak absorption coefficients of $5 \times 10^5$, $3 \times 10^5$, and $3 \times 10^5$ cm$_{-1}$ respectively. For comparison, CdTe has a peak absorption coefficient of $3 \times 10^4$ cm$_{-1}$ [87], and Si has a peak absorption coefficient of about $1 \times 10^5$ cm$_{-1}$ [88]. Another feature of interest is the absorption onset, which represents the energy at which the absorption begins to rise substantially. For this work, we define the absorption onset to be the point at which the absorption coefficient reaches 10$_4$ cm$_{-1}$, as in Ref. [62]. In order to be suitable for solar cells, a material should have a strong absorption onset in the range of 0.8–1.7 eV, bracketing the peak in the AM1.5 solar irradiance spectrum [10]. We calculate these absorption onsets for NaAsS$_2$ to be in the 1.87 to 2.20 eV range, depending on which of three directions, *a*, *b*, and *c*, is selected. Similarly, NaSbS$_2$ has an absorption onset range of 1.42 to 1.64 eV, and finally, NaSbSe$_2$ has absorption onsets ranging from 1.09 to 1.13 eV. The absorption onsets are slightly more than the band gaps of these compounds as is expected: NaAsS$_2$ (1.95 eV), NaSbS$_2$ (1.39 eV), and NaSbSe$_2$ (1.02 eV). The band gaps of other commonly used materials such as Si, CIGS, and CdTe are of comparable values: 1.14 eV [89], 1.0-1.7 eV [90], and 1.45 eV [91] respectively. Even though NaAsS$_2$ has a higher band gap, it could still be of some value in PV applications. These three compounds, all of the *C2/c* structure, could be tuned by alloying which could make them absorber layer candidates for photovoltaics. Alternatively, they could be used in properly designed tandem solar cells as one or more of the absorber materials.

## 4. Conclusion

We computed structural, electronic, thermoelectric, and optical properties of ternary chalcogenides NaAX$_2$, where A = As, Sb, Bi, and X = S, Se, Te using first principles computational methods. The structures were geometrically optimized, which allowed us to determine the lattice parameters of each compound. Our lattice parameters match well with available previous works, both experimental and theoretical, for $NaSbS_2$ and $NaSbSe_2$. We have also observed that the lattice constants $a$, $b$, and $c$ strictly increase as the ionic radius of the X atom increases for a fixed A atom. Additionally, we find that that the ratio of the van der Waals radius of A to the van der Waals radius of X can be used as an indicator for stability. After computing phonon dispersion curves, we predict only $NaAsS_2$, $NaSbS_2$, and $NaSbSe_2$ to be dynamically stable at zero temperature.

In addition, hybrid functionals were implemented to study electronic, thermoelectric, and optical properties. We observe band gaps ranging from 0.51 to 1.95 eV, with six compounds having predicted band gaps between 0.9 to 1.6 eV. These calculated electronic band gaps match well with experimental data where it is available. We find highly anisotropic electronic and optical properties as may be expected based on the layered crystal structure. Using COHP analysis and calculating -IpCOHP, we are able to explain trends in stability by observing that A-X bonding electronic states vary significantly with structure. Two thermoelectric properties, Seebeck coefficient and reduced power factor, were computed and found to have promising values. At a concentration of $10_{18}$ cm$_{-3}$, both p-type and n-type $NaAsS_2$ have very high Seebeck coefficients of over 500 $\mu$V/K at high temperatures. Additionally, n-type $NaSbS_2$ and $NaSbSe_2$ at 600 K have remarkable Seebeck coefficients of about 500 $\mu$V/K and 300 $\mu$V/K. Thus, we recommend $NaSbS_2$ and $NaSbSe_2$, as has been commented in earlier works, and our new prediction of $NaAsS_2$ as potential candidates for TE applications. In addition to dielectric functions, absorption coefficients were computed and found to have high values, from $10^4$ to $10^5$ cm$_{-1}$, in the visible range. These absorption coefficients, along with the band gaps, indicate that all three stable compounds, $NaAsS_2$, $NaSbS_2$, and $NaSbSe_2$, could be used in photovoltaic, especially in tandem solar cells, and thermoelectric applications.

## 5. Conflicts of Interest

There are no conflicts of interest.

## 6. Acknowledgements

The computing for this project was performed at the Ohio Supercomputer Center (OSC) [92]. We thank the University of Toledo's Research In Science and Engineering (RISE) Program for high school students [93]. We also thank the National Science Foundation grants CMMI 1629230 and 1629239 for funding this work.

**Table 1**

Structural parameters for the unit cell of each compound. Calculated lattice constants, *a, b, c, β,* and volume, *V*, are shown. All nine compounds are of the monoclinic lattice type, and of space group *C2/c* (#15). Parameters from previous works studying in the same space group are listed in parenthesis.

| Compound | *a* (Å) | *b* (Å) | *c* (Å) | β (°) | *V* (Å³) |
|---|---|---|---|---|---|
| NaAsS₂ | 7.78 | 8.17 | 6.47 | 121.76 | 87.32 |
| NaAsSe₂ | 7.95 | 8.52 | 6.80 | 121.08 | 98.54 |
| NaAsTe₂ | 8.32 | 9.09 | 7.32 | 120.85 | 118.81 |
| NaSbS₂[a] | 8.06 (8.23) | 8.10 (8.25) | 6.81 (6.84) | 123.61 (124.3) | 92.54 (95.9) |
| NaSbSe₂[b,c] | 8.24 (8.43) | 8.40 (8.72) | 7.12 (7.23) | 123.00 (123*, 120†) | 103.37 |
| NaSbTe₂ | 8.64 | 8.98 | 7.65 | 122.92 | 124.41 |
| NaBiS₂ | 7.95 | 7.95 | 7.04 | 124.29 | 91.81 |
| NaBiSe₂ | 8.28 | 8.29 | 7.31 | 124.18 | 103.79 |
| NaBiTe₂ | 8.80 | 8.96 | 7.79 | 124.03 | 127.34 |

[a] Experimental values from Ref. [21].

[b] Theoretical values from Ref. [20]. *The angle reported by Ref. [20], $\gamma$, is supplementary to $\beta$.

[c] Theoretical values from Ref. [70]. †The angle reported by Ref. [70], $\gamma$, is supplementary to $\beta$.

**Table 2**

Calculated electronic band gap for each compound. Also tabulated are the negative Integrated Projected Crystal Orbital Hamilton (-IpCOHP) Populations, for the cation-anion bonds in each compound. Energy integration up to the Fermi level yields IpCOHP, which serves as a bond-weighting indicator [56]. Theoretical results from previous works are given in parenthesis.

| Compound | Band gap (eV) | -IpCOHP (eV) |
|----------|---------------|--------------|
| $NaAsS_2$ | 1.95 | As-S: 3.39<br>Na-S: 0.72 |
| $NaAsSe_2$ | 1.52 | As-Se: 3.23<br>Na-Se: 0.74 |
| $NaAsTe_2$ | 0.97 | As-Te: 3.23<br>Na-Te: 0.96 |
| $NaSbS_2$ | 1.39<br>(1.22[a]) | Sb-S: 3.47<br>Na-S: 0.70 |
| $NaSbSe_2$ | 1.02<br>(0.843[b], 1.11[c]) | Sb-Se: 3.39<br>Na-Se: 0.77 |
| $NaSbTe_2$ | 0.51 | Sb-Te: 3.12<br>Na-Te: 0.93 |
| $NaBiS_2$ | 1.49 | Bi-S: 3.14<br>Na-S: 0.76 |
| $NaBiSe_2$ | 1.23 | Bi-Se: 2.99<br>Na-Se: 0.86 |
| $NaBiTe_2$ | 1.95 | Bi-Te: 2.72<br>Na-Te: 0.86 |

[a] Ref. [19].

[b] Ref. [20].

[c] Ref. [18].

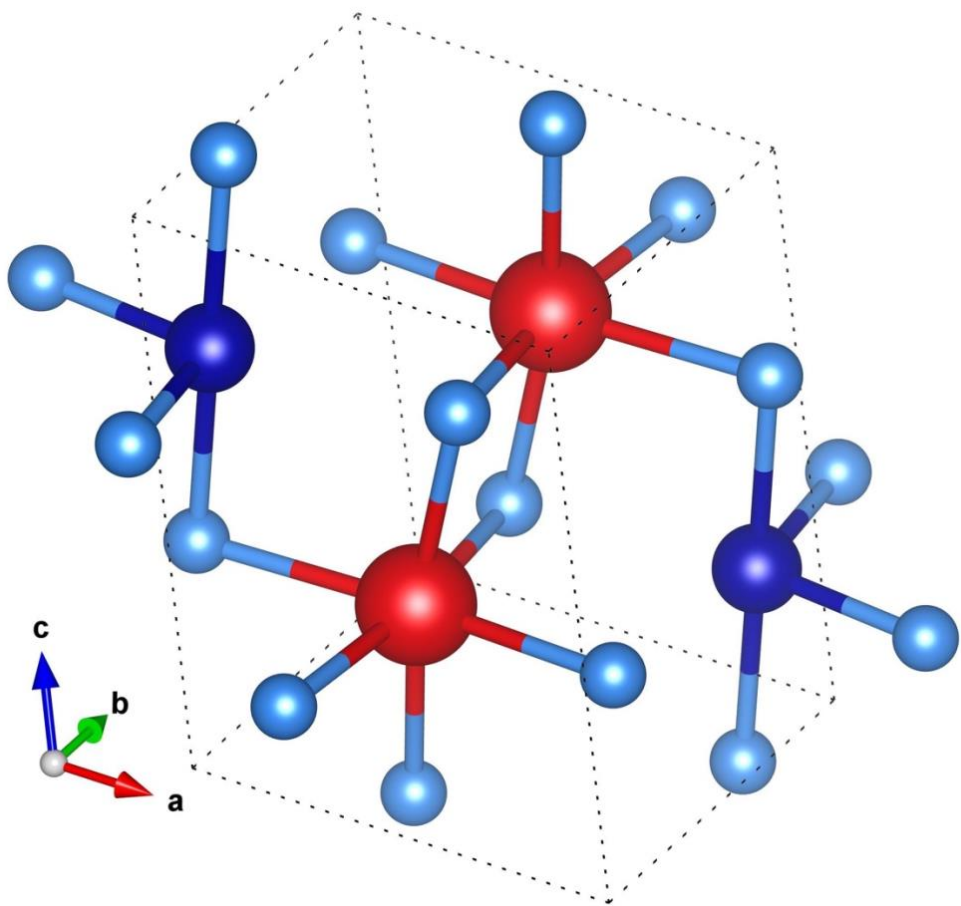

**Fig. 1.** The general unit cell of the form $NaAX_2$. The large red spheres represent Na atoms, the medium dark blue spheres represent A atoms, and the smaller light blue spheres represent X atoms. The visualization was obtained using VESTA [94].

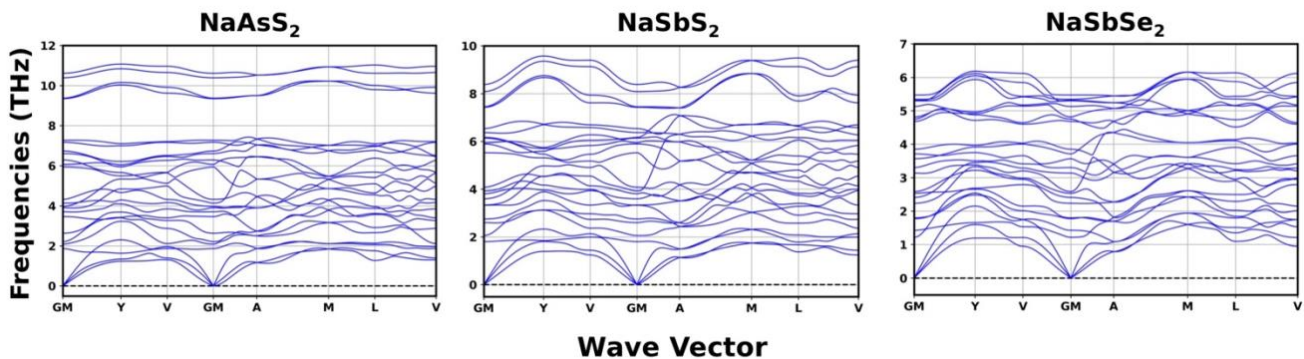

**Fig. 2.** Phonon dispersion curves for NaAsS$_2$, NaSbS$_2$, and NaSbSe$_2$, calculated using the harmonic approximation at 0 K. The dashed horizontal line represents zero frequency, separating positive (real) from negative (imaginary) frequencies. Only these three compounds show absence of imaginary frequencies, and are therefore predicted to be dynamically stable.

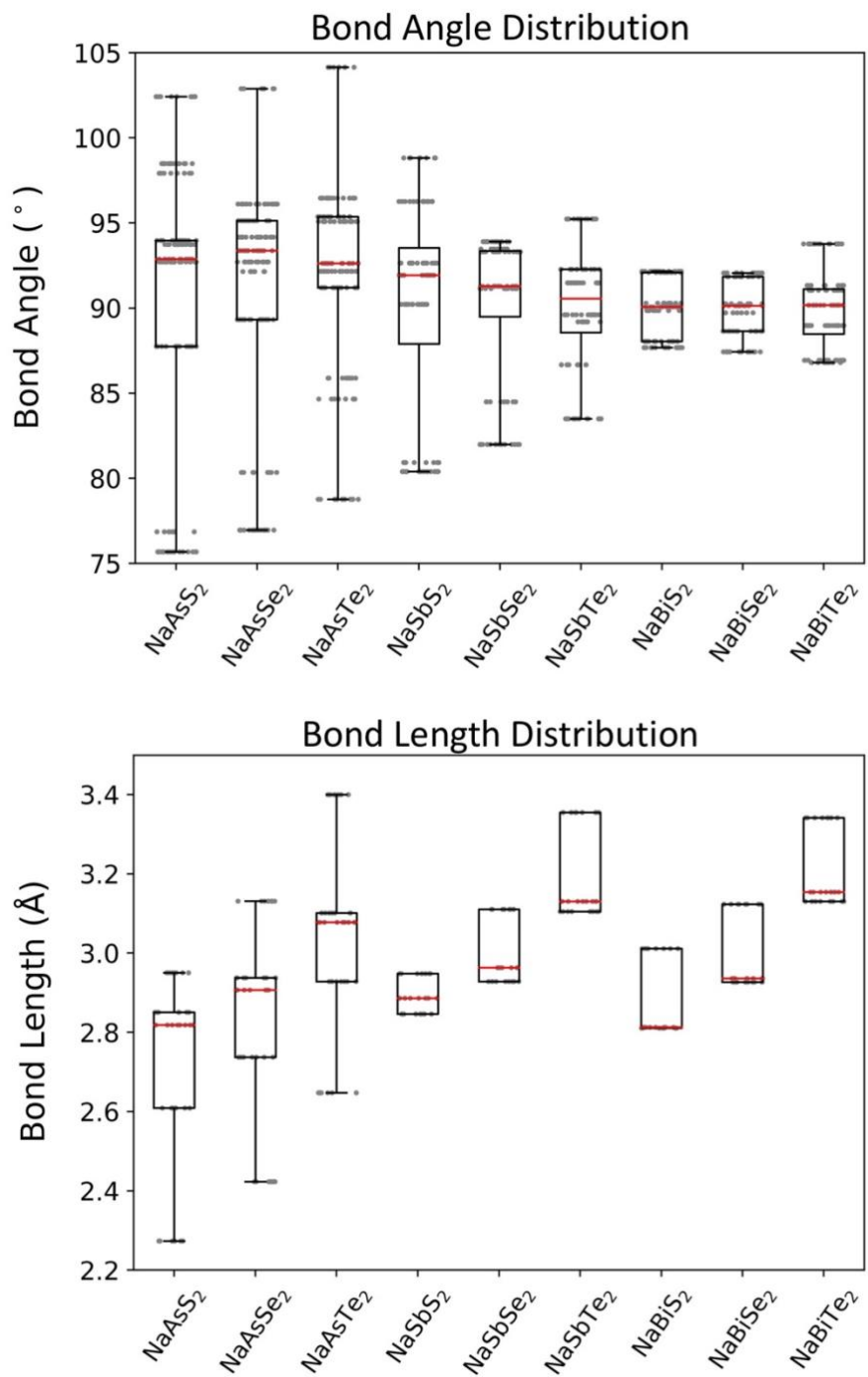

**Fig. 3.** The distribution of bond lengths and angles for all materials in this study. Each dot represents an individual bond length or angle within the unit cell. The median value of the lengths/angles is shown by the red lines. The boxes extend from the lower to upper quartile values of the data, whereas the whiskers extend to show the range of the data.

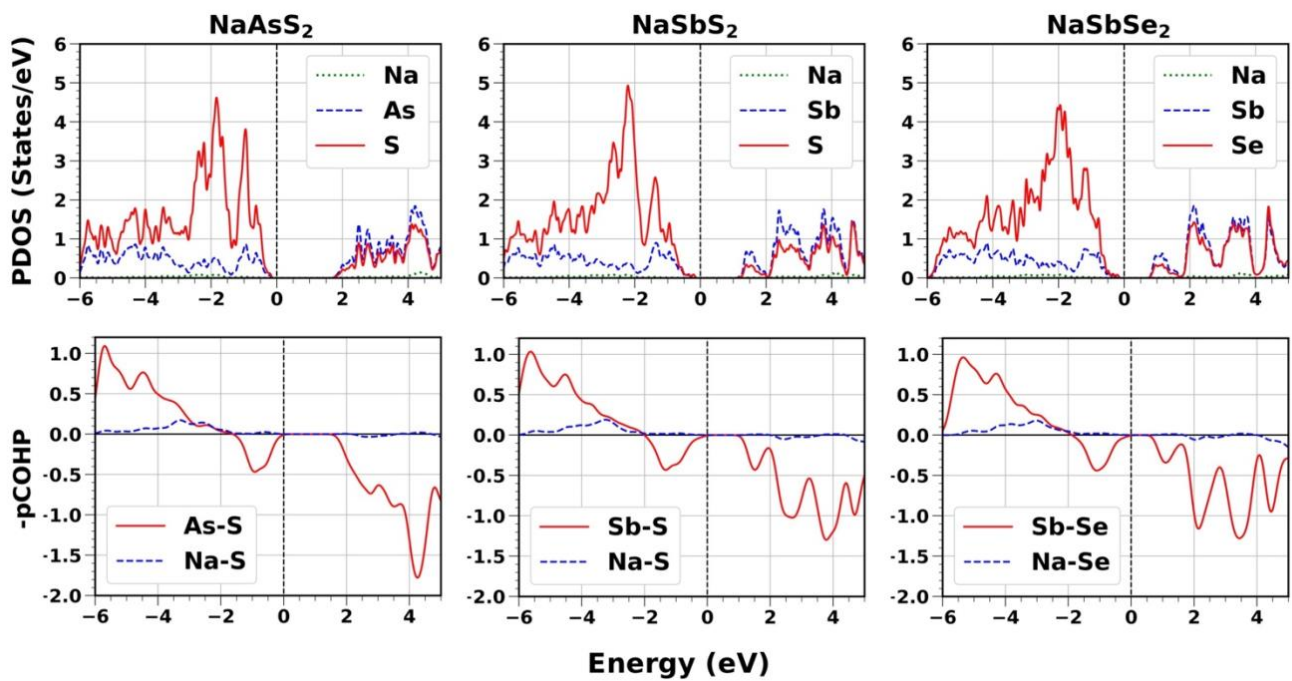

**Fig. 4.** Projected electronic density of states (PDOS) for $NaAsS_2$, $NaSbS_2$, and $NaSbSe_2$, computed through implementation of the HSE06 functional. Corresponding negative projected Crystal Orbital Hamiltonian Populations (-pCOHP) for $NaAsS_2$, $NaSbS_2$, and $NaSbSe_2$, separated into the two major bonding pairs between cation and anion are shown. Other bond populations are not shown here, as their magnitudes are insignificant. As plotted, positive values correspond with bonding states while negative values correspond with antibonding states. The Fermi energy is set to 0 eV.

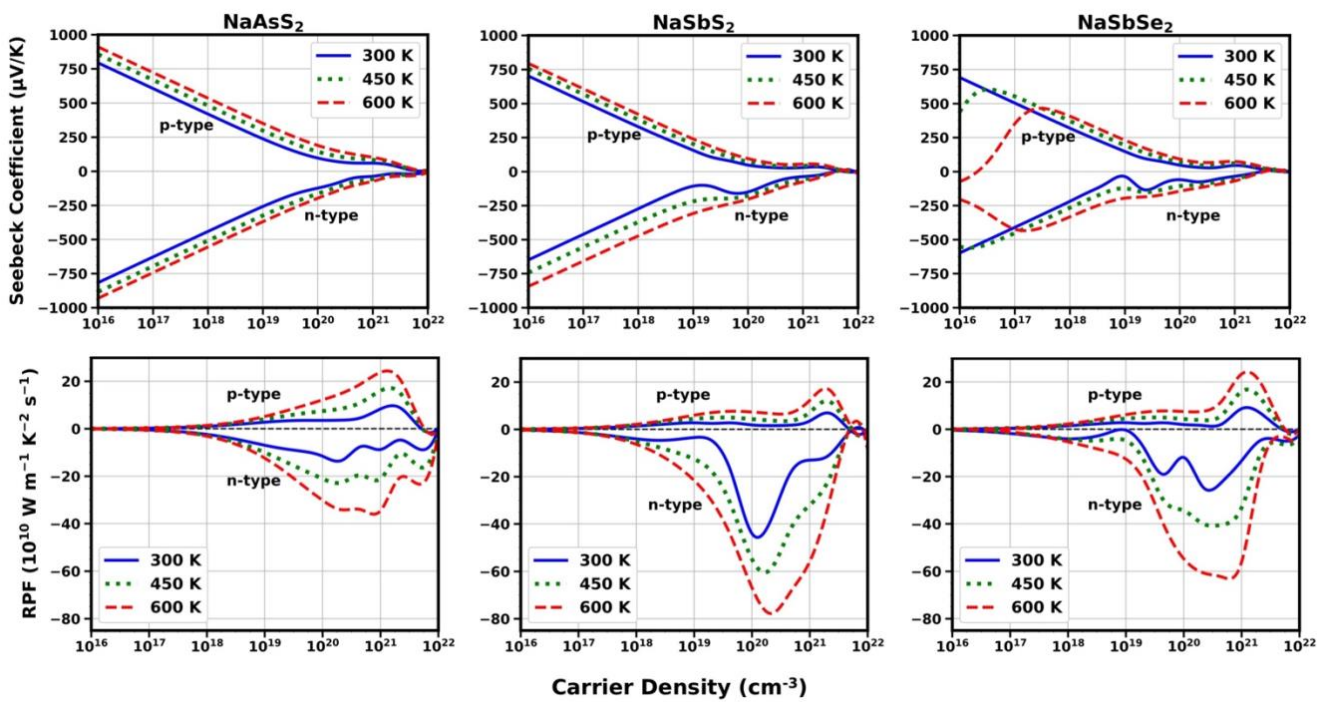

**Fig. 5.** Seebeck coefficients (in µV/K) of NaAsS$_2$, NaSbS$_2$, and NaSbSe$_2$ at various temperatures as a function of carrier density (cm$^{-3}$) for both p-type and n-type doping concentration. Reduced power factor (RPF), in (10$_{10}$ W m$_{-1}$ K$_{-2}$ s$_{-1}$), as a function of carrier density (cm$_{-3}$) at various temperatures for both p-type and n-type doping concentration are also shown. The temperatures plotted here are of interest for application in relevant thermoelectric devices.

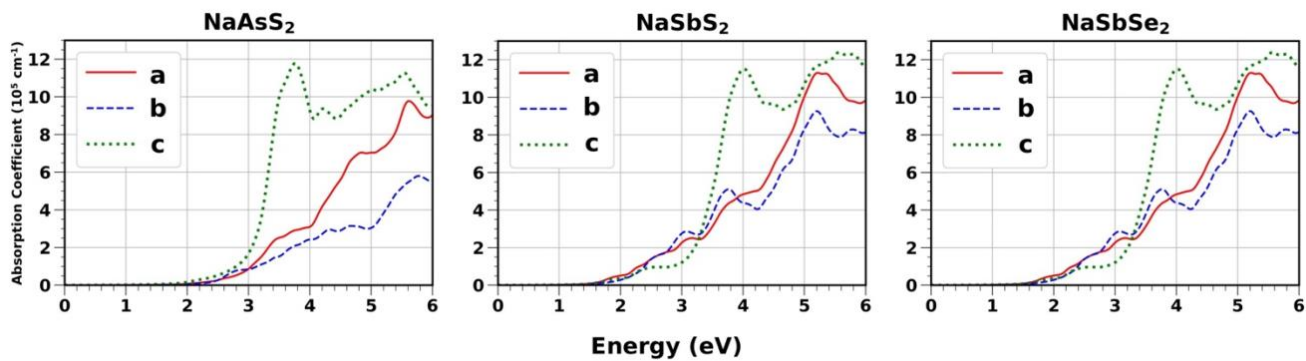

**Fig. 6.** Absorption coefficients $\alpha(E)$ of each compound, computed through implementation of the hybrid HSE06 functional. The three independent directional values are plotted to illustrate anisotropy arising from the monoclinic symmetry.

# References

[1] N. G. F. Reaver, et al., Imminence of peak in US coal production and overestimation of reserves, International Journal of Coal Geology. 131 (2014) 90.

[2] D. Bonnet, et al., Cadmium-telluride—Material for thin film solar cells, J. Mater. Res. 13 (1998) 2740.

[3] S. Moon, et al., Highly efficient single-junction GaAs thin-film solar cell on flexible substrate, Scientific Reports. 6 (2016) 30107.

[4] B. M. Başol, et al., Copper indium diselenide thin film solar cells fabricated on flexible foil substrates, Solar Energy Materials and Solar Cells. 29 (1993) 163.

[5] A. Kojima, et al., Organometal Halide Perovskites as Visible-Light Sensitizers for Photovoltaic Cells, J. Am. Chem. Soc. 131 (2009) 6050.

[6] A. Zakutayev, Design of nitride semiconductors for solar energy conversion, Journal of Materials Chemistry A. 4 (2016) 6742.

[7] B. K. Meyer, et al., Binary copper oxide semiconductors: From materials towards devices, physica status solidi (b). 249 (2012) 1487.

[8] S. Rühle, et al., All-Oxide Photovoltaics, The Journal of Physical Chemistry Letters. 3 (2012) 3755.

[9] H. Hoppe, et al., Organic solar cells: An overview, J. Mater. Res. 19 (2004) 1924.

[10] H. J. Goldsmid, et al., The use of semiconductors in thermoelectric refrigeration, Br. J. Appl. Phys. 5 (1954) 386.

[11] Z. H. Dughaish, Lead telluride as a thermoelectric material for thermoelectric power generation, Physica B: Condensed Matter. 322 (2002) 205.

[12] M. Christensen, et al., Thermoelectric clathrates of type I, Dalton Transactions. 39 (2010) 978.

[13] M. C. Schäfer, et al., Tin Clathrates with the Type II Structure, J. Am. Chem. Soc. 135 (2013) 1696.

[14] Y. Yin, et al., Recent advances in oxide thermoelectric materials and modules, Vacuum. 146 (2017) 356.

[15] K. Hoang, et al., Atomic Ordering and Gap Formation in Ag-Sb-Based Ternary Chalcogenides, Phys. Rev. Lett. 99 (2007) 156403.

[16] E. Peccerillo, et al., Copper–antimony and copper–bismuth chalcogenides—Research opportunities and review for solar photovoltaics, MRS Energy & Sustainability. 5 (2018) E13.

[17] W. W. W. Leung, et al., An experimental and theoretical study into NaSbS$_2$ as an emerging solar absorber, Journal of Materials Chemistry C. 7 (2019) 2059.

[18] C.-M. Dai, et al., NaSbSe$_2$ as a promising light-absorber semiconductor in solar cells: First-principles insights, APL Materials. 7 (2019) 081122.

[19] J. Sun, et al., Electronic Properties, Screening, and Efficient Carrier Transport in NaSbS$_2$, Physical Review Applied. 7 (2017) 024015.

[20] A. Putatunda, et al., Thermoelectric properties of layered NaSbSe$_2$, Journal of Physics: Condensed Matter. 30 (2018) 225501.

[21] J. Olivier-Fourcade, et al., Structure des composés NaSbS$_2\alpha$ et NaSbS$_2\beta$. Etude de l'influence de la paire electronique E de l'antimonine III dans la transition NaSbS$_2\alpha \rightarrow$ NaSbS$_2\beta$, Zeitschrift für anorganische und allgemeine Chemie. 446 (1978) 159.

[22] H. A. Graf, et al., Darstellung und Kristallstruktur von KSbS$_2$, Zeitschrift für anorganische und allgemeine Chemie. 414 (1975) 211.

[23] V. B. Lazarev, et al., Crystal growth of KSbSe$_2$, Mater. Res. Bull. 7 (1972) 417.

[24] B. Eisenmann, et al., Über Seleno- und Telluroarsenite, -antimonite und -bismutite, Zeitschrift für anorganische und allgemeine Chemie. 456 (1979) 87.

[25] V. A. Bazakutsa, et al., Photoelectric and Optical Properties of Thin Films of Ternary

Chalcogenides of the Form $Me_I SbX_{2 VI}$, Soviet Physics Journal. 18 (1975) 472.

[26] K. Hoang, et al., Atomic and electronic structures of I-V-VI$_2$ ternary chalcogenides, Journal of Science: Advanced Materials and Devices. 1 (2016) 51.

[27] S. U. Rahayu, et al., Sodium antimony sulfide (NaSbS$_2$): Turning an unexpected impurity into a promising, environmentally friendly novel solar absorber material, APL Materials. 4 (2016) 116103.

[28] G. Kresse, et al., Ab initio molecular dynamics for liquid metals, Physical Review B. 47 (1993) 558.

[29] G. Kresse, et al., Ab initio molecular-dynamics simulation of the liquid-metal--amorphous-semiconductor transition in germanium, Physical Review B. 49 (1994) 14251.

[30] G. Kresse, et al., Efficiency of ab-initio total energy calculations for metals and semiconductors using a plane-wave basis set, Computational Materials Science. 6 (1996) 15.

[31] G. Kresse, et al., Efficient iterative schemes for ab initio total-energy calculations using a plane-wave basis set, Physical Review B. 54 (1996) 11169.

[32] J. P. Perdew, et al., Atoms, molecules, solids, and surfaces: Applications of the generalized gradient approximation for exchange and correlation, Physical Review B. 46 (1992) 6671.

[33] P. E. Blöchl, Projector augmented-wave method, Physical Review B. 50 (1994) 17953.

[34] G. Kresse, et al., From ultrasoft pseudopotentials to the projector augmented-wave method, Physical Review B. 59 (1999) 1758.

[35] O. Jepson, et al., The electronic structure of h.c.p. Ytterbium, Solid State Commun. 9 (1971) 1763.

[36] A. Baldereschi, Mean-Value Point in the Brillouin Zone, Physical Review B. 7 (1973) 5212.

[37] B. B. Dumre, et al., Improved optoelectronic properties in CdSe$_x$Te$_{1-x}$ through controlled composition and short-range order, Solar Energy. 194 (2019) 742.

[38] D. J. Chadi, et al., Special Points in the Brillouin Zone, Physical Review B. 8 (1973) 5747.

[39] S. L. Cunningham, Special points in the two-dimensional Brillouin zone, Physical Review B. 10 (1974) 4988.

[40] H. J. Monkhorst, et al., Special points for Brillouin-zone integrations, Physical Review B. 13 (1976) 5188.

[41] K. Balasubramanian, et al., Vacancy-induced mechanical stabilization of cubic tungsten nitride, Physical Review B. 94 (2016)

[42] K. Balasubramanian, et al., Valence electron concentration as an indicator for mechanical properties in rocksalt structure nitrides, carbides and carbonitrides, Acta Materialia. 152 (2018) 175.

[43] J. A. Warner, et al., Ab initio calculations for properties of MAX phases Ti$_2$TlC, Zr$_2$TlC, and Hf$_2$TlC, Appl. Phys. Lett. 88 (2006)

[44] N. J. Szymanski, et al., Electronic and optical properties of vanadium oxides from first principles, Computational Materials Science. 146 (2018) 310.

[45] X. Zhou, et al., Mechanical properties and electronic structure of anti-ReO$_3$ structured cubic nitrides, M$_3$N, of d block transition metals M: An ab initio study, Journal of Alloys and Compounds. 595 (2014) 80.

[46] N. J. Szymanski, et al., Unconventional superconductivity in 3d rocksalt transition metal carbides, Journal of Materials Chemistry C. 7 (2019) 12619.

[47] M. J. Mehl, et al., The AFLOW Library of Crystallographic Prototypes: Part 1, Computational Materials Science. 136 (2017) S1.

[48] D. Hicks, et al., The AFLOW Library of Crystallographic Prototypes: Part 2, Computational Materials Science. 161 (2019) S1.

[49] B. Fultz, Vibrational thermodynamics of materials, Progress in Materials Science. 55 (2010) 247.

[50] N. J. Szymanski, et al., Prediction of improved magnetization and stability in Fe16N$_2$ through alloying, J. Appl. Phys. 126 (2019) 093903.

[51] A. Togo, et al., First principles phonon calculations in materials science, Scripta Materialia. 108

(2015) 1.

[52] A. V. Krukau, et al., Influence of the exchange screening parameter on the performance of screened hybrid functionals, The Journal of Chemical Physics. 125 (2006) 224106.

[53] R. Dronskowski, et al., Crystal orbital Hamilton populations (COHP): energy-resolved visualization of chemical bonding in solids based on density-functional calculations, The Journal of Physical Chemistry. 97 (1993) 8617.

[54] V. L. Deringer, et al., Crystal Orbital Hamilton Population (COHP) Analysis As Projected from Plane-Wave Basis Sets, The Journal of Physical Chemistry A. 115 (2011) 5461.

[55] S. Maintz, et al., Analytic projection from plane-wave and PAW wavefunctions and application to chemical-bonding analysis in solids, Journal of Computational Chemistry. 34 (2013) 2557.

[56] S. Maintz, et al., LOBSTER: A tool to extract chemical bonding from plane-wave based DFT, Journal of Computational Chemistry. 37 (2016) 1030.

[57] S. Maintz, et al., Efficient Rotation of Local Basis Functions Using Real Spherical Harmonics, Acta Physica Polonica B. 47 (2016)

[58] M. Gajdoš, et al., Linear optical properties in the projector-augmented wave methodology, Physical Review B. 73 (2006) 045112.

[59] R. Deng, et al., Optical and transport measurement and first-principles determination of the ScN band gap, Physical Review B. 91 (2015) 045104.

[60] Z. T. Y. Liu, et al., Transparency enhancement for $SrVO_3$ by $SrTiO_3$ mixing: A first-principles study, Computational Materials Science. 144 (2018) 139.

[61] E. Hecht, Optics, fourth ed., Addison-Wesley, Boston, 2002.

[62] N. J. Szymanski, et al., Dynamical stabilization in delafossite nitrides for solar energy conversion, Journal of Materials Chemistry A. 6 (2018) 20852.

[63] G. K. H. Madsen, et al., BoltzTraP. A code for calculating band-structure dependent quantities, Computer Physics Communications. 175 (2006) 67.

[64] A. Saini, et al., Effect of temperature dependent relaxation time of charge carriers on the thermoelectric properties of LiScX (X=C, Si, Ge) half-Heusler alloys, Journal of Alloys and Compounds. 806 (2019) 1536.

[65] P.-C. Wei, et al., Enhancing thermoelectric performance by Fermi level tuning and thermal conductivity degradation in $(Ge_{1-x}Bi_x)Te$ crystals, Scientific Reports. 9 (2019) 8616.

[66] T. Ren, et al., Manipulating Localized Vibrations of Interstitial Te for Ultra-High Thermoelectric Efficiency in p-Type Cu–In–Te Systems, ACS Applied Materials & Interfaces. 11 (2019) 32192.

[67] E. S. Toberer, et al., Zintl Chemistry for Designing High Efficiency Thermoelectric Materials, Chemistry of Materials. 22 (2010) 624.

[68] L. He, et al., Accuracy of generalized gradient approximation functionals for density-functional perturbation theory calculations, Physical Review B. 89 (2014) 064305.

[69] R. D. Shannon, Revised effective ionic radii and systematic studies of interatomic distances in halides and chalcogenides, Acta Crystallographica Section A. 32 (1976) 751.

[70] B. A. Aragaw, et al., Ion exchange-prepared $NaSbSe_2$ nanocrystals: electronic structure and photovoltaic properties of a new solar absorber material, RSC Adv. 7 (2017) 45470.

[71] S. Mukhopadhyay, et al., Optic phonons and anisotropic thermal conductivity in hexagonal $Ge_2Sb_2Te_5$, Scientific Reports. 6 (2016) 37076.

[72] M. K. Jana, et al., Crystalline Solids with Intrinsically Low Lattice Thermal Conductivity for Thermoelectric Energy Conversion, ACS Energy Letters. 3 (2018) 1315.

[73] A. Bondi, van der Waals Volumes and Radii, The Journal of Physical Chemistry. 68 (1964) 441.

[74] M. Mantina, et al., Consistent van der Waals Radii for the Whole Main Group, The Journal of Physical Chemistry A. 113 (2009) 5806.

[75] W. Shockley, et al., Detailed Balance Limit of Efficiency of p-n Junction Solar Cells, J. Appl.

Phys. 32 (1961) 510.

[76] S. Chen, et al., Crystal and electronic band structure of $Cu_2ZnSnX_4$ (X=S and Se) photovoltaic absorbers: First-principles insights, Appl. Phys. Lett. 94 (2009) 041903.

[77] D. B. Mitzi, et al., The path towards a high-performance solution-processed kesterite solar cell, Solar Energy Materials and Solar Cells. 95 (2011) 1421.

[78] Z. T. Y. Liu, et al., Electronic and bonding analysis of hardness in pyrite-type transition-metal pernitrides, Physical Review B. 90 (2014)

[79] H. S. Kim, et al., Relationship between thermoelectric figure of merit and energy conversion efficiency, Proceedings of the National Academy of Sciences. 112 (2015) 8205.

[80] M. D. Nielsen, et al., Lone pair electrons minimize lattice thermal conductivity, Energy & Environmental Science. 6 (2013) 570.

[81] X. Zhang, et al., Thermoelectric materials: Energy conversion between heat and electricity, Journal of Materiomics. 1 (2015) 92.

[82] M. M. A. Mahmoud, et al., Structural, stability and thermoelectric properties for the monoclinic phase of $NaSbS_2$ and $NaSbSe_2$: A theoretical investigation, The European Physical Journal B. 92 (2019) 214.

[83] D. J. Singh, Doping-dependent thermopower of PbTe from Boltzmann transport calculations, Physical Review B. 81 (2010) 195217.

[84] H. J. Goldsmid, Principles of thermoelectric devices, Br. J. Appl. Phys. 11 (1960) 209.

[85] S. Ju, et al., Revisiting PbTe to identify how thermal conductivity is really limited, Physical Review B. 97 (2018) 184305.

[86] M. K. Han, et al., Thermoelectric Properties of $Bi_2Te_3$: CuI and the Effect of Its Doping with Pb Atoms, Materials (Basel). 10 (2017)

[87] J. Rangel-Cardenas, et al., Optical Absorption Enhancement in CdTe Thin Films by Microstructuration of the Silicon Substrate, Materials (Basel). 10 (2017)

[88] C.-Y. Tsai, Absorption coefficients of silicon: A theoretical treatment, J. Appl. Phys. 123 (2018) 183103.

[89] B. Streetman, et al., Solid State Electronic Devices, 5th, Prentice Hall, New Jersey, 1999.

[90] T. Tinoco, et al., Phase Diagram and Optical Energy Gaps for $CuIn_yGa_{1-y}Se_2$ Alloys, physica status solidi (a). 124 (1991) 427.

[91] D. A. Jenny, et al., Semiconducting Cadmium Telluride, Phys. Rev. 96 (1954) 1190.

[92] Ohio Supercomputer Center, 1987. <http://osc.edu/ark:/19495/f5s1ph73>.

[93] University of Toledo's Research In Science & Engineering (RISE) Program, 2019. <https://www.utoledo.edu/research/pvic/rise.html>.

[94] K. Momma, et al., VESTA 3 for three-dimensional visualization of crystal, volumetric and morphology data, J. Appl. Crystallogr. 44 (2011) 1272.

# Electronic, optical and thermoelectric properties of sodium pnictogen chalcogenides: A first principles study

I. S. Khare[a,b], N. J. Szymanski[c], D. Gall[d], R. E. Irving[b],*

[a]Ottawa Hills High School, Ottawa Hills, OH 43606, USA

[b]Department of Physics and Astronomy, University of Toledo, Toledo, OH 43606, USA

[c]Department of Materials Science and Engineering, University of California, Berkeley, CA 94720, USA

[d]Department of Materials Science and Engineering, Rensselaer Polytechnic Institute, Troy, NY 12180, USA

*Corresponding Author: richard.irving@utoledo.edu

# Supplementary Material

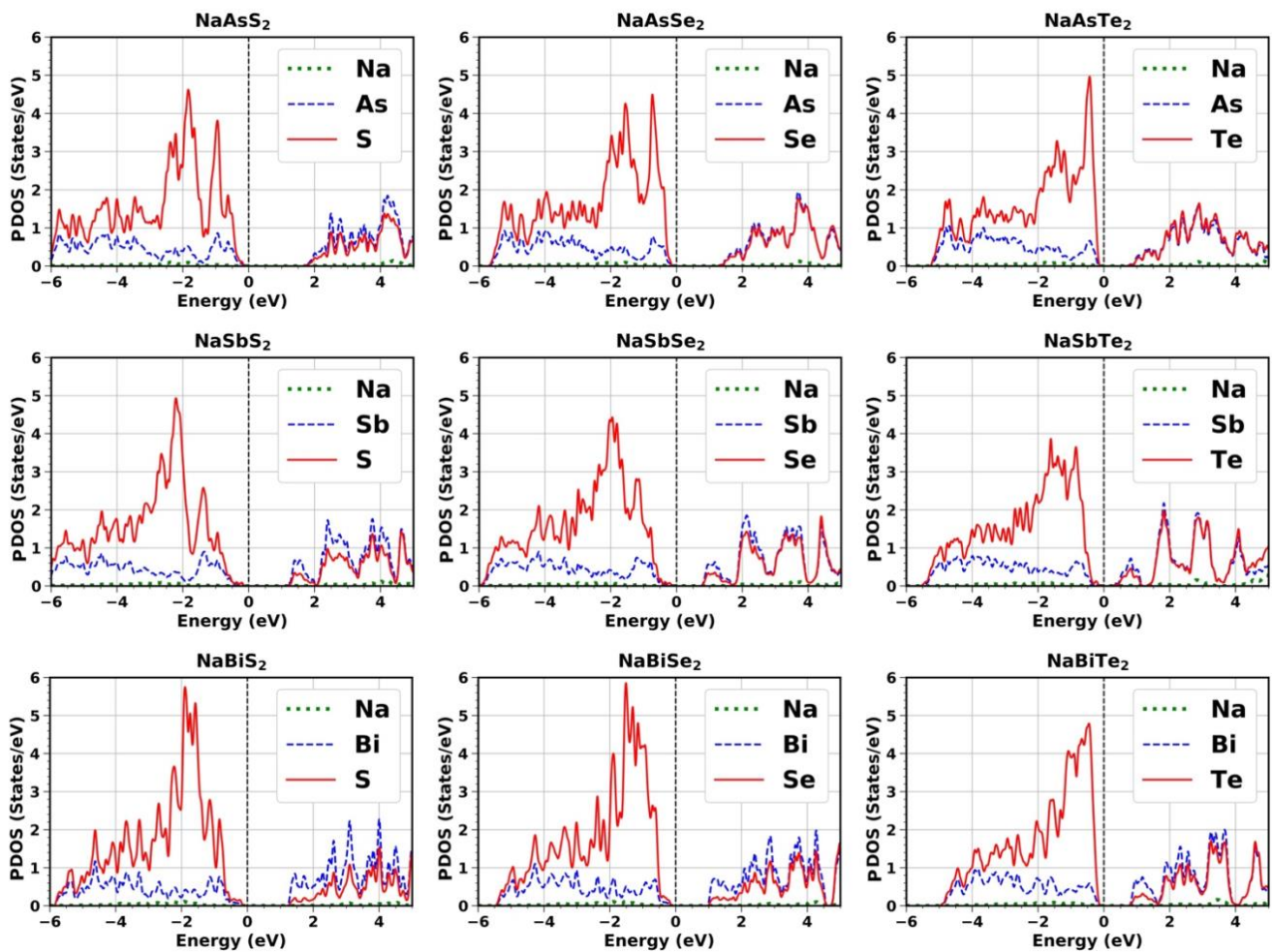

**Fig. S1:** Projected electronic density of states (PDOS), computed through implementation of the HSE06 functional. The Fermi energy is set to 0 eV.

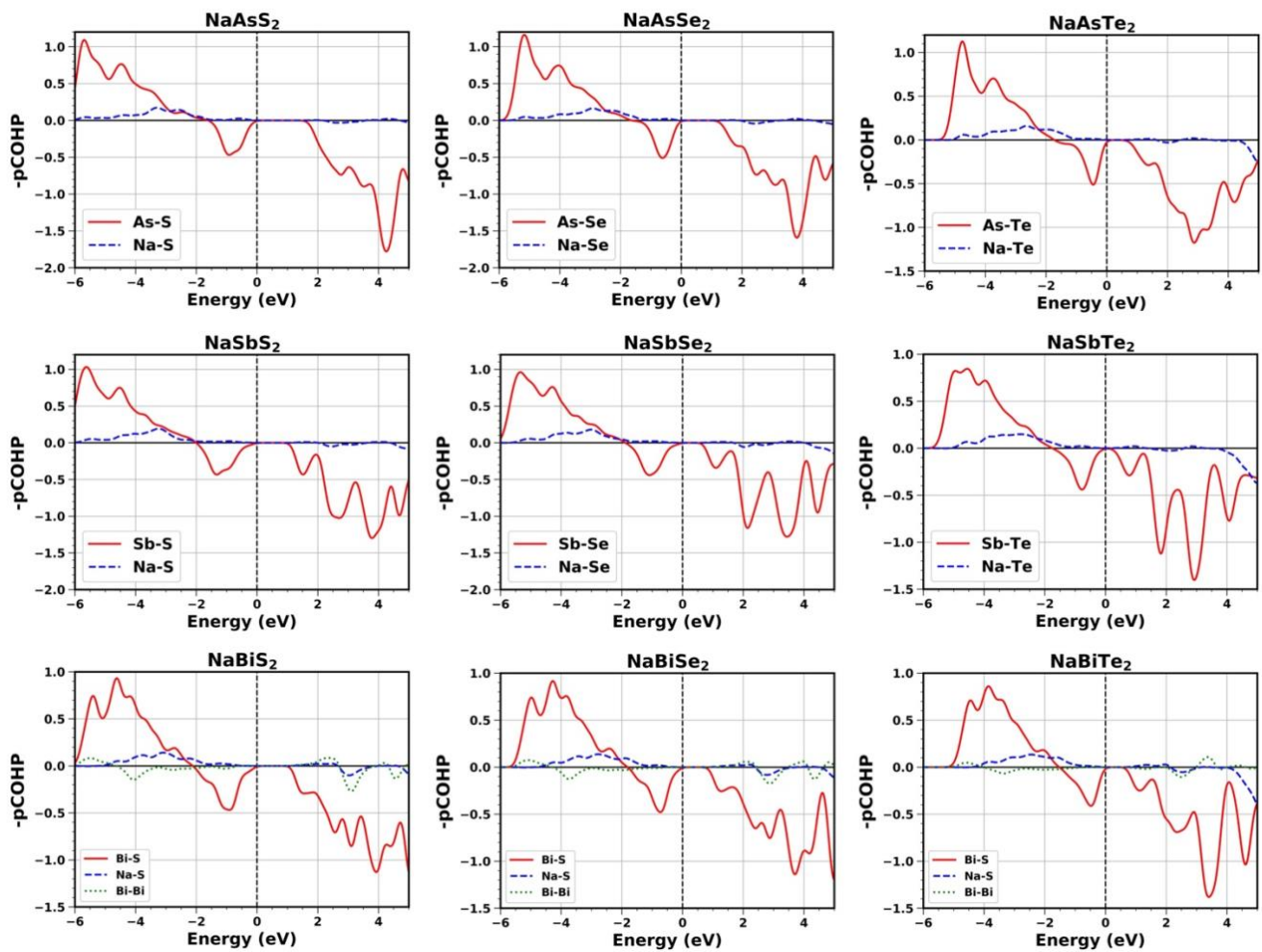

**Fig. S2:** Negative projected Crystal Orbital Hamiltonian Populations (-pCOHP), separated into the two major bonding pairs between cation and anion are plotted. Other bond populations are not shown here, as their magnitudes are insignificant. As plotted, positive values correspond with bonding states while negative values correspond with antibonding states. The Fermi energy is set to 0 eV.

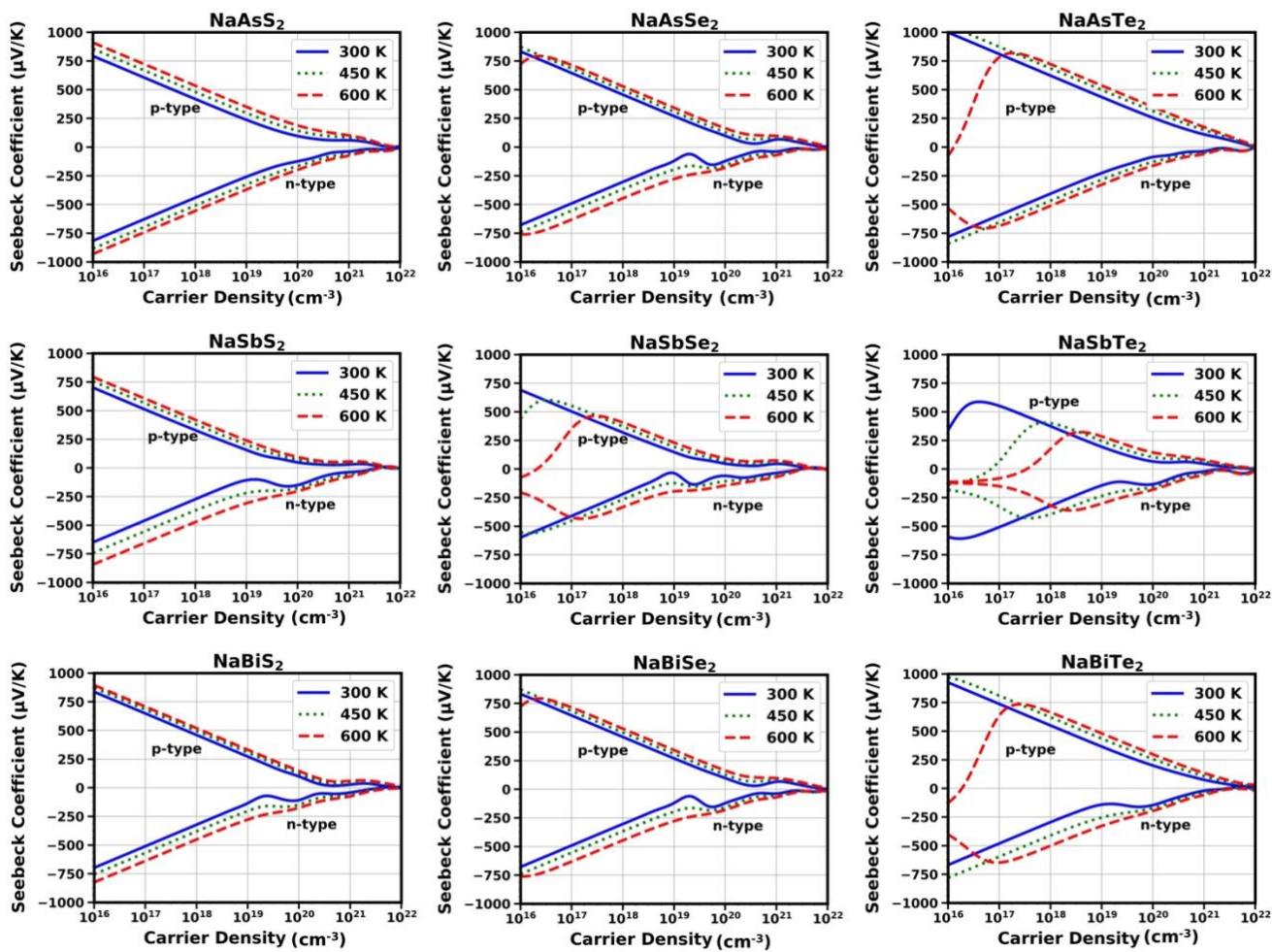

**Fig. S3:** Seebeck coefficients (in µV/K) as a function of carrier density (cm-3) for both p-type and n-type doping concentration at three different temperatures.

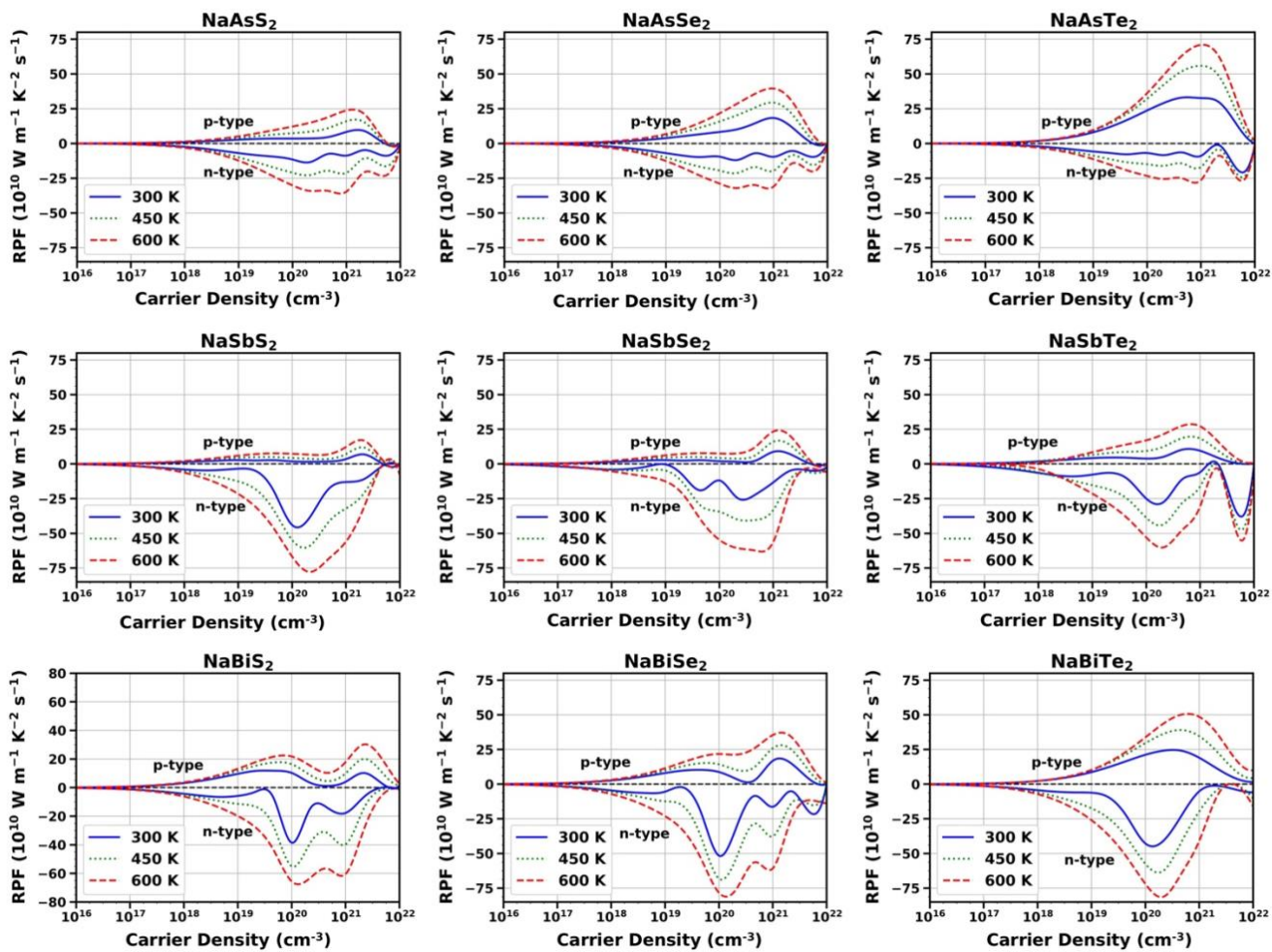

**Fig. S4:** Reduced power factor (RPF) in ($10^{10}$ W m$^{-1}$ K$_{-2}$ s$^{-1}$), as a function of carrier density (cm$_{-3}$) at various temperatures for both p-type and n-type doping concentration.

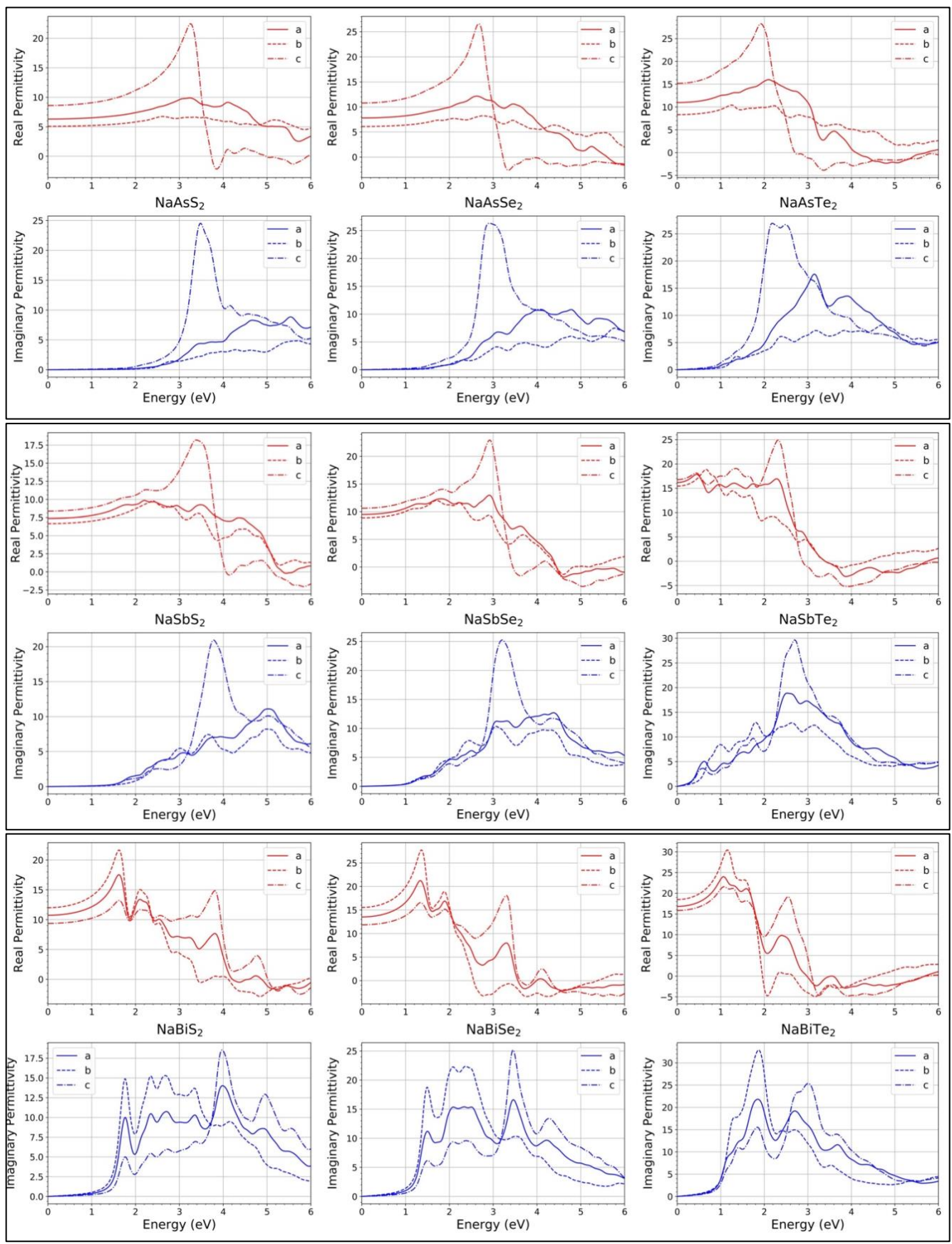

**Fig. S5:** The real and imaginary parts of the complex dielectric function, as calculated through implementation of the hybrid HSE06 functional. Due to the anisotropic nature of the materials, we have plotted the dielectric values in each direction according to the level of anisotropy within the compound.

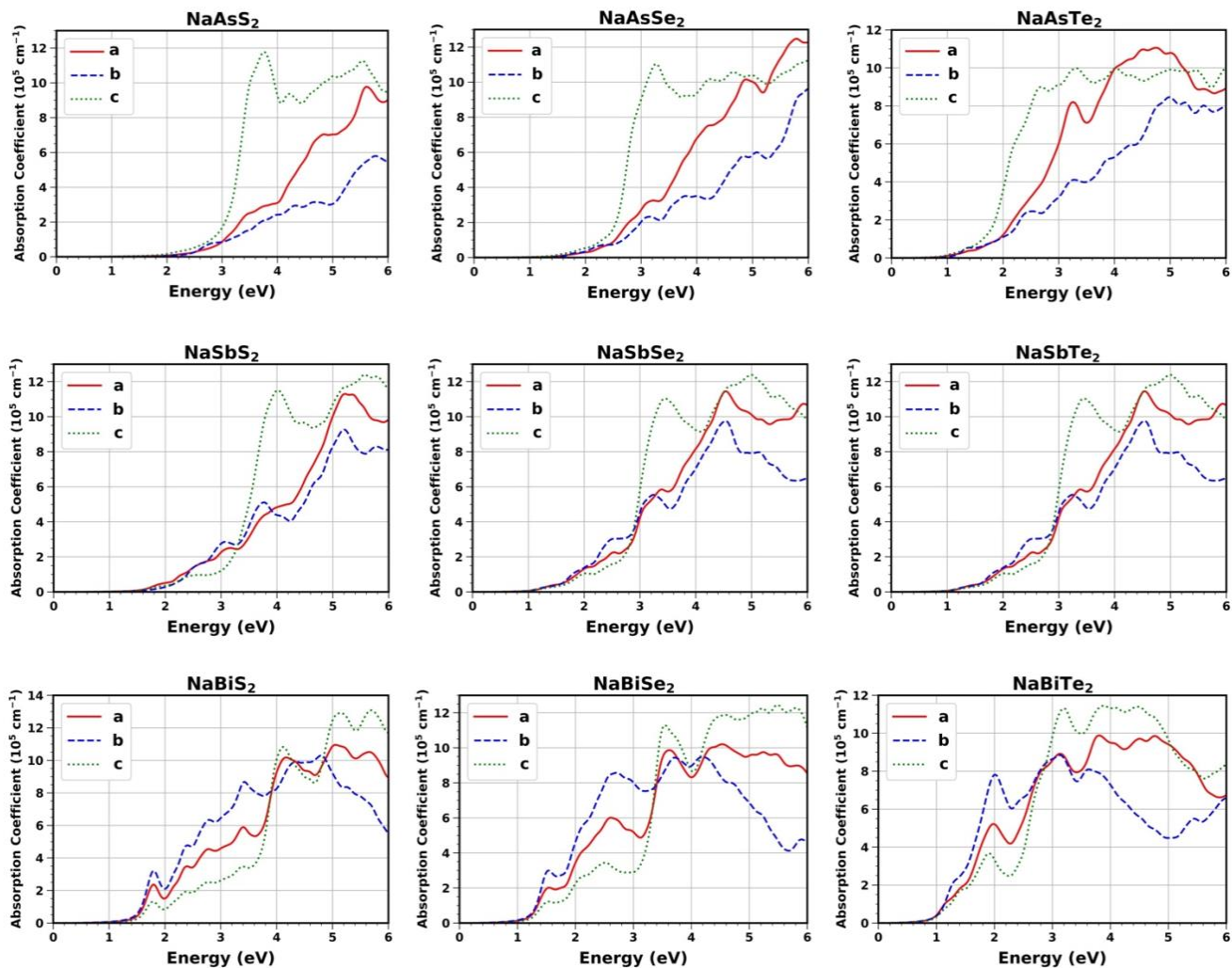

**Fig. S6:** Absorption coefficients $\alpha(E)$ of each compound, computed through implementation of the hybrid HSE06 functional. The three independent directional values are plotted to illustrate anisotropy arising from the monoclinic symmetry.